%% file: paraarxiv_final.tex
\documentclass[10pt,aps,prl,showpacs,twocolumn,amsmath,amssymb,superscriptaddress]{revtex4-1}
\usepackage{amssymb,amsthm,amsfonts,amstext}
\usepackage{color}
\usepackage{ifpdf}
\ifpdf 
  \usepackage[pdftex]{graphicx}
  \DeclareGraphicsExtensions{.pdf,.png,.jpg,.jpeg,.mps}
  \usepackage{pgf}
  \usepackage{tikz}
  \usepackage{hyperref}
\else 
  \usepackage{graphicx}
  \DeclareGraphicsExtensions{.eps,.bmp}
  \usepackage{pgf}
  \usepackage{tikz}
  \usepackage{pstricks}
\fi

\DeclareMathOperator{\trace}{tr}

\begin{document}

\title{An operational framework for nonlocality}

\author{Rodrigo Gallego}
\affiliation{ICFO-Institut de Ciencies Fotoniques, E-08860
Castelldefels, Barcelona, Spain}
\author{Lars Erik W\"{u}rflinger}
\affiliation{ICFO-Institut de Ciencies Fotoniques, E-08860
Castelldefels, Barcelona, Spain}
\author{Antonio Ac\'in}
\affiliation{ICFO-Institut de Ciencies Fotoniques, E-08860
Castelldefels, Barcelona, Spain} \affiliation{ICREA-Institucio
Catalana de Recerca i Estudis Avan\c{c}ats, Lluis Companys 23,
08010 Barcelona, Spain}
\author{Miguel Navascu\'es}
\affiliation{School of Physics, University of Bristol,
Bristol BS8 1TL, U.K.}

\begin{abstract}
Due to the importance of entanglement for quantum information
purposes, a framework has been developed for its characterization
and quantification as a resource based on the following
operational principle: entanglement among $N$ parties cannot be
created by local operations and classical communication, even when
$N-1$
parties collaborate. 
More recently, nonlocality has been identified as another
resource, alternative to entanglement and necessary for
device-independent quantum information
protocols. 
We introduce an operational framework for nonlocality based on a
similar principle: nonlocality among $N$ parties cannot be created
by local operations and allowed classical communication even when
$N-1$ parties collaborate. We then show that the standard
definition of
multipartite nonlocality, due to Svetlichny, 
is inconsistent with this operational approach: according to it,
genuine tripartite nonlocality could be created by two
collaborating parties. We finally discuss alternative definitions
for which consistency is recovered.
\end{abstract}

\maketitle


\textit{Introduction.} The fundamental importance of entanglement
in Quantum Information Science has driven a strong theoretical
effort devoted to its characterization, detection and
quantification. The resulting entanglement theory~\cite{horodecki}
has produced new mathematical tools, such as entanglement
witnesses or entanglement measures, which find application also
beyond the quantum information scenario for which they were
initially derived, e.g. in Condensed Matter Physics~\cite{fazio},
Quantum Thermodynamics~\cite{Popescu} or Biology~\cite{Sarovar}.

The first step when deriving this theoretical formalism consists
in identifying the relevant objects and set of operations, see
Table~\ref{table}. The relevant objects in the entanglement
scenario are quantum states in systems composed by $N$ observers,
labeled by $A_i$ with $i=1,\ldots,N$. The relevant set of
operations is the set of local operations and classical
communication (LOCC). The whole formalism then relies on the
following principle, which has a clear operational motivation:
{\em entanglement of a quantum state is a resource that cannot be
created by LOCC.} This implies that those states that can be
created by LOCC are not entangled. These states are called
separable and can be written as~\cite{horodecki}
\begin{equation}\label{sepst}
    \rho_{A_1\ldots A_N}=\sum_j p_j\rho_{A_1}^j\otimes\cdots\otimes\rho_{A_N}^j .
\end{equation}
In turn, those states that cannot be created by LOCC are entangled
and require a nonlocal quantum resource for the preparation. It is
easy to see that LOCC protocols map separable states into
separable states. Finally, entanglement witnesses are Hermitian
operators $W$ such that (i) $\trace(W\rho_S)\geq 0$ for all
separable states $\rho_S$ but (ii) there exist an entangled state
$\rho$ such that $\trace(W\rho)<0$. 

\begin{table}
\begin{center}
\begin{tabular}{|l|l|l|}
\hline
Resource& Objects & Operations\\
\hline
Entanglement &Quantum states & LOCC\\
Nonlocality & Joint Probability Distributions & WCCPI\\
\hline
\end{tabular}
\end{center}
\caption{Comparison of entanglement and non-locality from an
operational point of view. Once the basic ingredients of the
theory have been identified, an operational framework is based on
the following principle: the resource contained in the states
cannot increase under the set of operations.} \label{table}
\end{table}

The picture becomes richer when considering intermediate cases
where only some of the $N$ parties share entangled states. For
simplicity  we restrict our considerations in what follows to
three parties. Consider an entangled state in which only two
parties, say $A_2$ and $A_3$ are entangled. The corresponding
state is called biseparable and can be written as
\begin{equation}\label{bisepst}
    \rho_{A_1A_2A_3}=\sum_j p_j\rho_{A_1}^j\otimes\rho_{A_2A_3}^j .
\end{equation}
This state is not genuine 3-partite entangled, as for its LOCC
creation, it suffices that two of the parties act together.
Similarly as above, (i) LOCC protocols where $A_2$ and $A_3$ act
together map biseparable states into biseparable states along the
bipartition $A_1-A_2A_3$ and (ii) these states do not violate any
entanglement witness along this partition. 

Recently, a novel paradigm in Quantum Information Theory has been
introduced: device-independent quantum information
processing~\cite{diqkd,dirng}. Here, the main goal is to achieve
an information task without making any assumptions about the
internal working of the devices used in the protocol. The
device-independent property of these applications make them
appealing, both from a theoretical and implementation viewpoint.
In this scenario, the main objects are correlated systems
distributed among $N$ observers. Each observer $i$ can introduce a
classical input $x_i$ in his system, which produces a classical
output $a_i$. The system is just seen as a black box and no
assumption is made about the internal process producing the output
given the input, except that it cannot contradict quantum theory.
The correlations among the input/output processes in each system
are described by the joint conditional probability distribution
$P(a_1,\ldots,a_N|x_1,\ldots,x_N)$ of getting results
$a_1,\ldots,a_N$ when using the inputs $x_1,\ldots,x_N$. The main
reason why device-independent applications are possible in the
quantum regime is because of the existence of nonlocal quantum
correlations. Intuitively, since these correlations do not have a
classical analogue, they allow for novel protocols with no
classical counterpart.

The advent of device-independent applications leads to the
identification of nonlocality as a novel quantum information
resource, alternative to entanglement. Despite the only known way
of getting nonlocal quantum correlations among different observers
is by measuring entangled states, it is a well-established fact
that entanglement and nonlocality represent inequivalent quantum
properties~\cite{ADGL,Scarani}. Now, similar to what happened for
entanglement, it would be desirable to have an operational
framework for the characterization and quantification of
nonlocality as a resource. This is precisely the main motivation
for this work.


\textit{The operational framework.} In a similar way as it is done
for entanglement, the first step consists in identifying the
relevant objects and set of operations, see Table~\ref{table}. The
relevant objects are the joint probability distributions
$P(a_1,\ldots,a_N|x_1,\ldots,x_N)$. The corresponding set of
operations should include local processing of the classical inputs
and outputs. Communication is allowed only if it takes place
before the inputs are known, otherwise it can be used to create
nonlocal correlations. Such communication can be used either to
generate shared randomness or to announce the outcomes of a
sequence of measurements prior to the realization of the nonlocal
experiment. A general protocol in the nonlocality scenario would
thus begin with a \emph{preparation phase}, where one of the
parties would measure its system and broadcast the measurement
outcome. On the basis of that result, a second party would measure
its system, etc. At the end of the preparation phase, the parties
exchange some shared randomness and announce that they are ready
for the nonlocal experiment. Communication between them is
forbidden from this point on. The second step is the
\emph{measurement phase}, where each party is given an input, or
question, and they compute the outcome or answer by using the
correlations resulting from the \emph{preparation phase} and by
processing the obtained classical information at will. The last
process is commonly referred to as
`wirings'. 
Thus, in the nonlocality framework, the set of relevant operations
is Wirings \& Classical Communication Prior to the Inputs (WCCPI).

Once these two ingredients are identified, it is straightforward
to obtain an operational definition of nonlocality:  {\em
nonlocality of correlations $P(a_1,\ldots,a_N|x_1,\ldots,x_N)$ is
a resource that cannot be created by WCCPI.}

Not surprisingly, this operational definition leads to the
standard definition of nonlocality due to Bell~\cite{Bell} when
considering $N$ distant parties. Indeed, it is easy to see that
the correlations that can be created by WCCPI have the form, see
Eq.~\eqref{sepst},
\begin{multline}
\label{loccorr}
    P_\mathrm{L}(a_1,\ldots,a_n|x_1,\ldots,x_N)=\\
    \sum_\lambda p(\lambda)P_1(a_1|x_1,\lambda)\ldots P_1(a_n|x_n,\lambda) ,
\end{multline}
in which the local maps $P_i(a_i|x_i,\lambda)$ produce the
classical output $a_i$ depending on the input $x_i$ and a shared
classical random variable $\lambda$. All correlations that admit a
decomposition~\eqref{loccorr} are local, while they are nonlocal
otherwise. WCCPI protocols map local correlations into local
correlations. Finally, if we collect all the probabilities
$P(a_1,\ldots,a_N|x_1,\ldots,x_N)$ into a vector $\vec P$, any
Bell inequality can be seen as a vector of real coefficients $\vec
c$ such that (i) $\vec c\cdot \vec P_\mathrm{L}\geq 0$ for all
local correlations $\vec P_\mathrm{L}$ but (ii) there exist
correlations $\vec P$ such that $\vec c\cdot \vec P<0$.

As for entanglement, the next step is to characterize genuine
multipartite nonlocality. This question has already been studied
and the standard definition of genuine multipartite nonlocalty is
due to Svetlichny~\cite{svetlichny}. We restrict our
considerations again to three parties and the partition
$A_1-A_2A_3$, without loss of generality. According to Svetlichny,
correlations that can be written as, see Eq.~\eqref{bisepst},
\begin{multline}
\label{svcorr}
 P(a_1,a_2,a_3|x_1,x_2,x_3)=\\ \sum_\lambda p(\lambda)P_1(a_1|x_1,\lambda) P_{23}(a_2,a_3|x_2,x_3,\lambda)
\end{multline}
do not contain any genuine tripartite nonlocality, as there is a
local decomposition when parties $A_2$ and $A_3$ are together.
Correlations admitting a decomposition like~\eqref{svcorr} are
named in what follows bilocal (BL). As it happened for
entanglement and LOCC, it is expected that under WCCPI protocols
along the partition $A_1-A_2A_3$, bilocal correlations are mapped
into bilocal correlations. Consequently, no bipartite Bell
inequality between $A_1$ and $A_2A_3$ can be violated. Remarkably,
we prove here that this intuition is incorrect. This implies that
the standard definition of genuine multipartite nonlocality, given
by \eqref{svcorr}, is inconsistent with the operational approach.
In the following we show examples of the inconsistencies and also
provide and discuss alternative definitions of genuine
multipartite nonlocality that are consistent with our operational
framework.

First, we show the inconsistencies of the definition of BL by
providing correlations that (i) have a decomposition of the form
\eqref{svcorr} and (ii) become nonlocal along the partition
$A_1-A_2A_3$ when a WCCPI protocol, where $A_2$ and $A_3$
collaborate, is implemented. An example of these correlations with
a quantum realization can be established in the simplest scenario
consisting of two measurements of two outcomes for each of the
three observers. The measurements are performed on the quantum
state $|GHZ\rangle = \frac{1}{\sqrt{2}}(|000\rangle +
|111\rangle)$. The first observer, $A_1$ measures $\sigma_z$ and
$\sigma_x$, labeled by $x_1=0$ and $x_1=1$ respectively. $A_2$
measures $\sigma_z$ and $\sigma_x$, labeled by $x_2=0$ and $x_2=1$
respectively.  $A_3$ measures $\frac{\sigma_z+\sigma_x}{\sqrt{2}}$
and $\frac{\sigma_z-\sigma_x}{\sqrt{2}}$, labeled by $x_3=0$ and
$x_3=1$ respectively. These correlations have a decomposition of
the form \eqref{svcorr}. This can be easily computed by a linear
program, see Ref. \cite{Gallego}.

Let us now see how these tripartite collaborations can be mapped
into nonlocal bipartite correlations along the partition
$A_1-A_2A_3$ with an WCCPI protocol in which $A_2$ and $A_3$
collaborate. The protocol works as follows (see also
Figure~\ref{simple_wiring}.b): the first observer $A_1$ obtains
the output by using trivially his share of the tripartite box.
$A_2$ and $A_3$ collaborate by using the output obtained by $A_2$
as input for $A_3$. The resulting tripartite probability
distribution reads $P(a_1,a_2,a_3|x_1,x_2,x_3=a_2)$. The final
output is $A_3$'s output, so that the final probability
distribution is
$P(a_1,a_3|x_1,x_2)=\sum_{a_2}P(a_1,a_2,a_3|x_1,x_2,a_2)$. This
bipartite probability distribution $P(a_1,a_3|x_1,x_2)$ does not
have a local model. This can be verified by calculating the value
of the Clauser-Horne-Shimony-Holt (CHSH) polynomial
\begin{equation}\label{chsh}
\beta = C(0,0) + C(0,1) + C(1,1) -C(1,0),
\end{equation}
with $C(x_1,x_2)=P(a_1=a_3|x_1,x_2)-P(a_1\neq a_3|x_1,x_2)$. The
value obtained is $\beta=\frac{3}{\sqrt{2}}\approx 2.12$. Local
correlations fulfill $\beta \leq 2$, thus we conclude that the
correlations are nonlocal along the partition $A_1-A_2A_3$.

Alternatively, one can assess the inconsistency of Svetlichny's
definition by noting that our tripartite example behaves
non-locally if one of the parties broadcasts its measurement
outcomes before the nonlocal experiment takes place.
Indeed, suppose that, prior to the experiment, $A_2$ measures
$x_2=1$. If the result is $a_2=1$, $A_1$ and $A_3$ are projected
onto the distribution $P'(a_1,a_3|x_1,x_3)\equiv
P(a_1,a_3|x_1,x_3,x_2=1,a_2=1)$. On the contrary, if $A_2$ reads
$a_2=-1$, $A_1$ receives the order of inverting her measurement
outcomes for measurement $x_1=1$, and so the system is projected
again into $P'(a_1,a_3|x_1,x_3)=P(-a_1,a_3|x_1,x_3,x_2=1,a_2=-1)$.
It can be checked that the new bipartite distribution
$P'(a_1,a_3|x_1,x_3)$ violates the CHSH inequality maximally
($\beta=2\sqrt{2}\approx 2.82$).

\begin{figure}\label{simple_wiring}
 \centering
 \includegraphics[width=0.48 \textwidth]{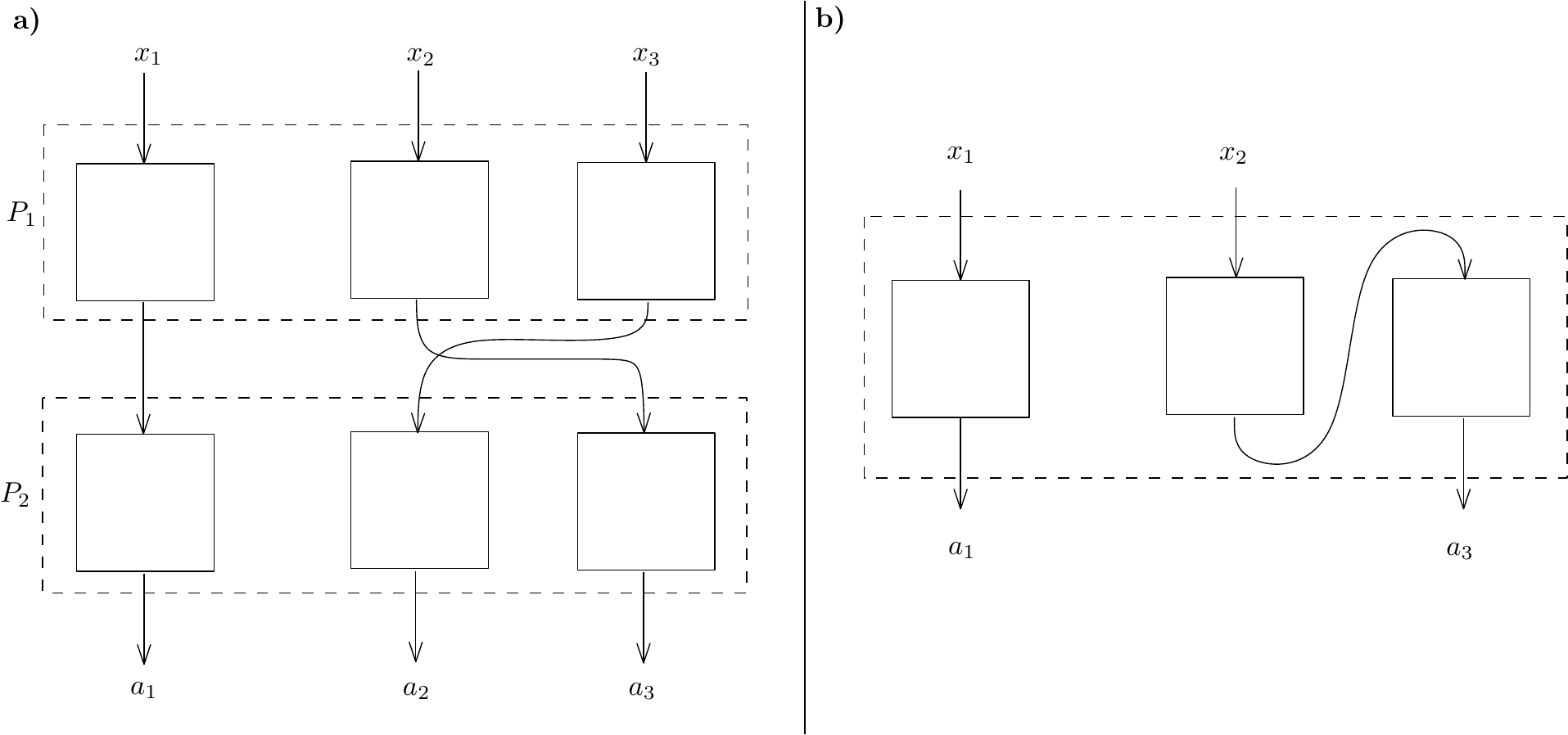}
 \caption{WCCPI protocols. a) Example of a WCCPI protocol
consistent with the definition of genuine nonlocality provided by
Eq.~\eqref{svcorr} (BL). The resulting probability distribution
$P(a_1,a_2,a_3|x_1,x_2,x_3)$ does not violate any Bell inequality
along the bipartition $A_1-A_2A_3$. However, this may be no longer
true for WCCPI protocols where inputs depend on outputs produced
by the collaborating parties. b) WCCPI protocol in which the input
of party $A_3$ is the output of $A_2$. As shown in the text, this
protocol can map tripartite probability distributions with a
bilocal decomposition as in \eqref{svcorr} into bipartite
distributions $P(a_1,a_3|x_1,x_2)$ that violate a Bell inequality.
This proves that this type of WCCPI protocols is not compatible
with the definition of genuine multipartite
nonlocality~\eqref{svcorr}.} \label{simple_wiring}
\end{figure}


As mentioned above, the existence of these correlations implies
that the standard definition of genuine multipartite
nonlocality~\eqref{svcorr} is inconsistent with our operational
approach, as it would imply that genuine tripartite nonlocality
could be created by WCCPI when two parties collaborate
\cite{note}. Thus, the concept of genuine multipartite nonlocality
is not correctly captured by Eq.~\eqref{svcorr}.

Now, the natural question is whether there are definitions of
bilocality which do not suffer from these inconsistencies. Or in
other words, whether one can find an analogous version of equation
~\eqref{bisepst} in the context of nonlocality consistent with the
operational framework established by WCCPI. Before moving into
that, it is worth understanding why the previous Bell violation is
possible even if the correlations seem to have a proper local
decomposition. The main reason is that no structure is imposed on
the joint terms $P_{23}(a_2,a_3|x_2,x_3,\lambda)$; in particular,
these terms may be incompatible with the no-signalling principle,
e.g. $ P_{23}(a_2|x_2,x_3,\lambda)\neq
P_{23}(a_2|x_2,x'_3,\lambda)$ for some $\lambda$. Further, if no
structure is imposed on $P_{23}(a_2,a_3|x_2,x_3,\lambda)$ the
decomposition ~\eqref{svcorr} may include terms which display both
signalling from $A_2$ to $A_3$ and from $A_3$ to $A_2$; that is,
the outcome probability distribution of $A_2$ depends on $A_3$'s
input and viceversa. Hence, a decomposition including these terms
cannot be considered a physical description of the situation in
which one of the observers measures first. This is crucial in our
previous example.

In our protocol the output of one of the parties is used as the
input of the other party, or it is broadcast prior to the
nonlocality experience. This implicitly assumes a temporal order
in the measurements which is inconsistent with such
decompositions. Indeed, all the examples of distributions of the
form~\eqref{svcorr} leading to a Bell violation under WCCPI have
to be such that the bilocal decomposition requires  terms
displaying signalling in both directions. Whether the converse is
true, that is, whether every decomposition with such terms can be
mapped via LOCC into a nonlocal one is an interesting open
question. We come back to this point below.

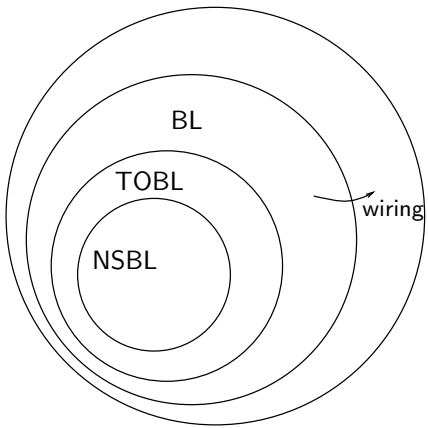
\begin{figure}
 \centering
 \input{sets.pspdftex}
 \caption{Representation of
no-signalling, time-ordered and general bilocal correlations. We
prove here that the set of no-signaling bilocal correlations
(NSBL) is strictly contained in the set of time-ordered bilocal
correlations (TOBL). The set TOBL is closed under wirings in the
sense that LOCC protocols involving two collaborating parties, say
$A_2$ and $A_3$, map TOBL correlations to correlations which are
local with respect to the partition $A_1-A_2 A_3$. The set of
general bilocal correlations (BL) however, contains correlations
that can be mapped by LOCC protocols to correlations that violate
a Bell inequality in the bipartition $A_1-A_2 A_3$.
}
 \label{fig:sets}
\end{figure}

It is now clear that tripartite correlations with bilocal
models~\eqref{svcorr} such that all the terms
$P_{23}(a_2,a_3|x_2,x_3,\lambda)$ satisfy the no-signalling
principle, i.e. marginal distributions on $A_2$ ($A_3$) do not
depend on the input by $A_3$ ($A_2$) for all $\lambda$, are
consistent with our operational framework. We name these
correlations no-signalling bilocal (NSBL). They are operationally
understood as correlations obtained by collaborating parties
sharing no-signalling resources. This definition however is too
restrictive, as it excludes correlations obtained by protocols in
which the collaborating parties communicate, which is perfectly
valid within our framework. Indeed, we show next that NSBL
correlations do not define the largest set of correlations
compatible with our framework, see Figure \ref{fig:sets}. 

Consider instead the set of tripartite no-signalling correlations
(see the Appendix for the corresponding $N$-party generalization)
that can be decomposed as
\begin{equation}
\begin{aligned}
\label{eq:tobl}
  P(a_1 a_2 a_3|x_1 x_2 x_3)
  &= \sum_\lambda p_\lambda P(a_1|x_1, \lambda) P_{2\rightarrow 3}(a_2 a_3|x_2 x_3,\lambda) \\
  &= \sum_\lambda p_\lambda P(a_1|x_1, \lambda) P_{2\leftarrow 3}(a_2
  a_3|x_2  x_3,\lambda)
\end{aligned}
\end{equation}
with the distributions $P_{2\rightarrow 3}$ and $P_{2\leftarrow
3}$ obeying the conditions
\begin{align}
 \label{eq:timeordered1}
  &P_{2 \rightarrow 3}(a_2|x_2,\lambda) = \sum_{a_3} P_{2 \rightarrow 3}(a_2 a_3|x_2 x_3,\lambda),\\
  \label{eq:timeordered2}
  &P_{2 \leftarrow 3}(a_3|x_3,\lambda) = \sum_{a_2} P_{2 \leftarrow 3}(a_2 a_3|x_2 x_3,\lambda).
\end{align}
We say that these correlations admit a \emph{time-ordered bilocal
(TOBL)} model. As can be seen from relations
\eqref{eq:timeordered1} and \eqref{eq:timeordered2} we impose the
distributions $P_{2\rightarrow 3}$ and $P_{2\leftarrow 3}$ to
allow for signaling at most in one direction, indicated by the
arrow. Decomposition \eqref{eq:tobl} has also been considered
in~\cite{Pironio}, and has a clear operational meaning:
$P(a_1a_2a_3|x_1x_2x_3)$ can be simulated by a classical random
variable $\lambda$ with probability distribution $p_\lambda$
distributed between parts $A_1$ and the composite system $A_2A_3$.
Using this variable, $A_1$ generates the output according to the
distribution $P(a_1|x_1, \lambda)$; on the other side, the two
outputs $a_2, a_3$ are generated using either $P_{2 \rightarrow
3}(a_2 a_3|x_2 x_3,\lambda)$ or $P_{2 \leftarrow 3}(a_2 a_3|x_2
x_3,\lambda)$, depending on which of the inputs $x_2$ or $x_3$ is
used first. Likewise, the bipartite distributions generated by a
prior postselection on, say, outcome $\tilde{a}_2$ of measurement
$\tilde{x}_2$ of system $B$ can be simulated locally by
$P(a_1|x_1,\lambda),P_{2\to
3}(a_3|x_3,x_2=\tilde{x}_2,a_2=\tilde{a}_2,\lambda)$ after an
appropriate modification of the weights $\{p_\lambda\}$ in eq.
(\ref{eq:tobl}).

TOBL correlations are thus consistent with our operational point
of view, as any WCCPI protocol along the partition $A_1-A_2A_3$
maps TOBL correlations~\eqref{eq:tobl} into probability
distributions with a local model along this
partition~\cite{Gallego}. Interestingly, this set is strictly
larger than the set of NSBL correlations (see Figure
\ref{fig:sets}). 

To prove this result, we consider the `Guess Your Neighbor's
Input' (GYNI) polynomial~\cite{Almeida2010}
\begin{multline}
\label{GYNI} \beta_{\mathrm{GYNI}}(P(a_1,a_2,a_3|x_1,x_2,x_3))=\\
P(000|000)+P(110|011)+P(011|101)+P(101|110).
\end{multline}
The maximum of this quantity over the set of probabilities having
a NSBL decomposition is equal to $1$, that is
$\beta_{\mathrm{GYNI}}(P \in \mathrm{NSBL})\leq 1$. In fact,
consider the terms in the NSBL decomposition
$P_1(a_1|x_1,\lambda)P_{23}^\mathrm{NS}(a_2,a_3|x_2,x_3,\lambda)$.
Without loss of optimality, one can restrict the analysis to
correlations where $P_1(a_1|x_1,\lambda)$ is deterministic, say
$P_1(0|0,\lambda)=P_1(0|1,\lambda)$. Thus, the GYNI polynomial for
this set of probabilities satisfies
\begin{equation}
\begin{aligned}
 &\beta_\mathrm{GYNI}(P(a_1,a_2,a_3|x_1,x_2,x_3,\lambda))\\
&=P_{23}^\mathrm{NS}(0,0|0,0,\lambda)+P_{23}^\mathrm{NS}(1,1|0,1,\lambda) \\
&\leq P_2(0|0,\lambda)+P_2(1|0,\lambda) \leq 1
\end{aligned}
\end{equation}
with
$P_2(a_2|x_2,\lambda)=\sum_{a_3}P_{23}^\mathrm{NS}(a_2,a_3|x_2,x_3,\lambda)$
being a well-defined distribution due to the no-signalling
constraints. One can easily check that the bound holds for any
other deterministic choice of $P_1(0|0,\lambda)$ and
$P_1(0|1,\lambda)$. As the NSBL decomposition is a convex mixture
of these points, the GYNI polynomial is also bounded by $1$. Note,
however, that in Ref. \cite{Gallego} it is shown that there is a
set of probabilities in TOBL obtaining larger values of the GYNI
polynomial. Hence, the set of NSBL is strictly contained in TOBL.


\textit{Conclusions.} We have introduced a novel framework for the
characterization of nonlocality which has an operational
motivation and captures the role of nonlocality as a resource for
device-independent quantum information processing. In spite of its
simplicity, the framework questions the current understanding of
genuine multipartite nonlocality, as the standard definition
adopted by the community is inconsistent with it. Similar
conclusions are reached from another perspective
in~\cite{Pironio}. We provide alternative frameworks where
consistency is recovered. The main open question is now to
identify the largest set of correlations that remain consistent
under WCCPI protocols when some of the parties collaborate. We
conjecture that TOBL correlations constitute such a set and,
therefore, that for any bilocal model requiring two-way signalling
terms there is a valid WCCPI protocol
detecting its inconsistency. 

\paragraph*{Acknowledgements}
We thank Stefano Pironio for discussions and, in particular, for
suggesting us the GHZ correlations leading to inconsistencies with
\eqref{svcorr}. This work was supported by the ERC starting grant
PERCENT, the CHIST-ERA DIQIP project, the European EU FP7
Q-Essence and QCS projects, the Spanish FIS2010-14830 project,
CatalunyaCaixa and the Templeton Foundation.






\newpage

\textit{Appendix: TOBL model for arbitrary number of parties.}
Suppose that $M+N$ parties share a no-signalling set of
correlations $P(a_1,...,a_{M+N}|x_1,...,x_{M+N})$. We are
interested in which restrictions we should enforce over such a
distribution in order to make sure that it cannot be used to
violate a bipartite Bell inequality when parties $1,...,M$ and
$M+1,...,M+N$ group together, even when several of such boxes are
initially distributed.

One possibility is to demand the new bipartite object to behave as
a generic classical bipartite device would. Viewed as bipartite,
the distribution $P(a_1,...,a_{M+N}|x_1,...,x_{M+N})$ is such that
it allows each of the two virtual parties (call them Alice and
Bob) to perform sequential measurements on their subsystems. If we
assume that the outcomes Alice and Bob observe are generated by a
classical machine, it follows that
$P(a_1,...,a_{M+N}|x_1,...,x_{M+N})$ can be written as:
\begin{equation}
    P(a_1,...,a_{M+N}|x_1,...,x_{M+N})=\sum_\lambda p_\lambda
P_A^\lambda \cdot P_B^\lambda,
\end{equation}
where we can regard each $P_A^\lambda$ as a collection of
probability distributions
\begin{equation}
    P^\lambda_{\sigma(1)\to...\to\sigma(M)}(a_1,...,a_M|x_1,...,x_M),
\end{equation}
one for each possible permutation $\sigma$ of the $M$ physical
parties. Here $\sigma(1)\to...\to\sigma(M)$ would indicate the
process in which the first party to measure is $\sigma(1)$,
followed by $\sigma(2)$, etc.

If, during a communication protocol, Alice must measure, say,
$x_3$, she only has to choose an arbitrary permutation $\sigma$,
with $\sigma(1)=3$ and then generate $a_3$ according to the
probability distribution
$P^\lambda_{\sigma(1)\to...\to\sigma(M)}(a_1,...,a_M|x_1,...,x_M)$.
If, at some time later, she needs to simulate the measurement of
$x_1$ and $\sigma(2)\not=1$, she would thus have to find a new
permutation $\sigma'$, with $\sigma'(1)=3,\sigma'(2)=1$, and
generate $a_1$ from the conditional probability distribution
$P^\lambda_{\sigma'(1)\to...\to\sigma'(M)}(a_1,a_2,a_4,...,a_M|x_1,...,x_M,a_3)$.
By consistency, for any pair of permutations $\sigma^1,\sigma^2$
such that $\sigma^1(j)=\sigma^2(j)$, for all $j\in \{1,...,m\}$,
such distributions need to satisfy the condition:
\begin{eqnarray}
&\sum_{a>m}P^\lambda_{\sigma^1(1)\to ...\to\sigma^1(M)}(a_1,...,a_{M}|x_1...x_M)=\nonumber\\
&=\sum_{a>m}P^\lambda_{\sigma^2(1)\to
...\to\sigma^2(M)}(a_1,...,a_{M}|x_1...x_M),
\end{eqnarray}
where $\sum_{a>m}$ denotes the sum over all variables
$a_{\sigma(j)}$ with $j>m$. The same considerations apply for
$P_B^\lambda$.

Local postselections on a prior sequence of Alice's and Bob's
outcomes would imply changing the probabilities $p_\lambda$, but
otherwise can be simulated in a similar fashion.

Putting everything together, we have that WCCPI operations over a
set of (possibly different) boxes generate bipartite classical
correlations if each box $P(a_1,...,a_{N+M}$$|x_1,...,x_{M+N})$,
distributed along the partition $1...M|M+1...M+N$, admits a
decomposition of the form
\begin{widetext}
\begin{eqnarray}
&P(a_1,...,a_{M+N}|x_1,...,x_{M+N})=\sum_{\lambda}p_\lambda P^\lambda_{\sigma(1)\to...\to\sigma(M)}(a_{\sigma(1)},...,_{\sigma(M)}|x_{\sigma(1)}...x_{\sigma(M)})\cdot\nonumber\\
&\cdot \tilde{P}^\lambda_{\sigma'(M+1)\to
...\to\sigma'(M+N)}(a_{\sigma'(M+1)},...,_{\sigma'(N+M)}|x_{\sigma'(N+1)}...x_{\sigma'(N+M)}).
\end{eqnarray}
\end{widetext}
The reader can check that in the tripartite case the above
description reduces to the TOBL definition given in the main text.

\end{document}

%% file: sets.pspdftex
\begin{picture}(0,0)%
\includegraphics{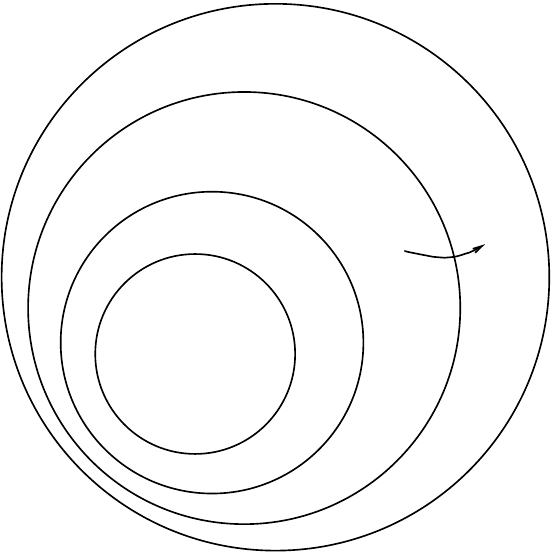}%
\end{picture}%
\setlength{\unitlength}{4144sp}%
\begingroup\makeatletter\ifx\SetFigFont\undefined%
\gdef\SetFigFont#1#2#3#4#5{%
  \reset@font\fontsize{#1}{#2pt}%
  \fontfamily{#3}\fontseries{#4}\fontshape{#5}%
  \selectfont}%
\fi\endgroup%
\begin{picture}(2520,2514)(6411,-2311)
\put(7069,-900){\makebox(0,0)[lb]{\smash{{\SetFigFont{10}{12.0}{\rmdefault}{\mddefault}{\updefault}{\color[rgb]{0,0,0}{\sf TOBL}}%
}}}}
\put(7405,-530){\makebox(0,0)[lb]{\smash{{\SetFigFont{10}{12.0}{\rmdefault}{\mddefault}{\updefault}{\color[rgb]{0,0,0}{\sf BL}}%
}}}}
\put(6933,-1371){\makebox(0,0)[lb]{\smash{{\SetFigFont{10}{12.0}{\rmdefault}{\mddefault}{\updefault}{\color[rgb]{0,0,0}{\sf NSBL}}%
}}}}
\put(8551,-1051){\makebox(0,0)[lb]{\smash{{\SetFigFont{9}{10.8}{\rmdefault}{\mddefault}{\updefault}{\color[rgb]{0,0,0}{\sf wiring}}%
}}}}
\end{picture}%